\documentclass[fleqn,usenatbib]{mnras}

\usepackage{newtxtext,newtxmath}
\usepackage{multicol}

\usepackage[T1]{fontenc}
\usepackage{hyperref}
\usepackage{gensymb}

\DeclareRobustCommand{\VAN}[3]{#2}
\let\VANthebibliography\thebibliography
\def\thebibliography{\DeclareRobustCommand{\VAN}[3]{##3}\VANthebibliography}

\usepackage{graphicx}	%
\usepackage{amsmath}	%
\usepackage{placeins}
\usepackage{xcolor}
\usepackage[normalem]{ulem}

\newcommand{\Rom}[1]
    {\uppercase\expandafter{\romannumeral #1\relax}}

\title[Detectability of microlensed GWs]{Detectability of microlensed gravitational waves}

\author[Simon M. C. Yeung et al.]{
Simon M. C. Yeung,$^{1,2}$\thanks{E-mail: yeungm@uwm.edu}
Mark H. Y. Cheung,$^{3}$
Eungwang Seo,$^{1,4}$
Joseph A. J. Gais,$^{1}$
Otto A. Hannuksela,$^{1}$
\newauthor
and Tjonnie G. F. Li$^{1,5,6}$
\\
$^{1}$Department of Physics, The Chinese University of Hong Kong, Shatin, N.T., Hong Kong.\\
$^{2}$University of Wisconsin-Milwaukee, Milwaukee, WI 53202, USA \\
$^{3}$William H. Miller III Department of Physics and Astronomy, Johns Hopkins University, 3400 North Charles Street, Baltimore, Maryland, 21218, USA. \\
$^{4}$SUPA, School of Physics and Astronomy, University of Glasgow, Glasgow G12 8QQ, United Kingdom.\\
$^{5}$Institute for Theoretical Physics, KU Leuven, Celestijnenlaan 200D, B-3001 Leuven, Belgium.\\
$^{6}$Department of Electrical Engineering (ESAT), KU Leuven, Kasteelpark Arenberg 10, B-3001 Leuven, Belgium.\\
}

\date{\today}

\pubyear{2021}

\begin{document}
\label{firstpage}
\pagerange{\pageref{firstpage}--\pageref{lastpage}}
\maketitle

\begin{abstract}
Gravitational lensing describes the bending of the trajectories of light and gravitational waves due to the gravitational potential of a massive object. 
Strong lensing by galaxies can create multiple images with different overall amplifications, arrival times, and image types. 
If, furthermore, the gravitational wave encounters a star along its trajectory, microlensing will take place. 
Previously, it has been shown that the effects of microlenses on strongly-lensed type-\Rom{1} images could be negligible in practice, at least in the low magnification regime.
In this work, we study the same effect on type-\Rom{2} strongly-lensed images by computing the microlensing amplification factor.
As opposed to being magnified, type-\Rom{2} images are typically demagnified. 
Moreover, microlensing on top of type-\Rom{2} images induces larger mismatches with un-microlensed waveforms than type-\Rom{1} images. 
These results are broadly consistent with recent literature and serve to confirm the findings. 
In addition, we investigate the possibility of detecting and analysing microlensed signals through Bayesian parameter estimation with an isolated point mass lens template, which has been adopted in recent parameter estimation literature.
In particular, we simulate gravitational waves microlensed by a microlens embedded in a galaxy potential near moderately magnified type-\Rom{1} and \Rom{2} macroimages, with variable lens masses, source parameters and macromagnifcations. 
Generally, an isolated point mass model could be used as an effective template to detect a type-\Rom{2} microlensed image but not for type-\Rom{1} images, demonstrating the necessity for more realistic microlensing search templates.
\end{abstract}

\section{Introduction}

Gravitational lensing takes place for both electromagnetic (EM) waves and gravitational waves (GWs).
We have many examples of EM lensing observations, while the lensing of GWs have long be predicted~\citep{Ohanian:1974ys, Thorne:1982cv, Wang:1996as}. 
Recently, searches for GW lensing using LIGO and Virgo data have also started~\citep{Hannuksela:2019kle, Li:2019osa, McIsaac:2019use, Pang:2020qow, Dai:2020tpj, Liu:2020par, o3alensing}, and many of the detection methodologies have been outlined in recent years, utilizing matched filtering~\citep{Li:2019osa,McIsaac:2019use} and Bayesian inference~\citep{Cao:2014oaa,Dai:2020tpj,Liu:2020par,Hannuksela:2019kle,Wang:2021,Pang:2020qow,Lo:2021nae,Janquart:2021nus}. 
If observed, lensed GWs could enable studies spanning the domains of, for example, fundamental physics and tests of general relativity~\citep{Baker:2016reh, Fan:2016swi, Mukherjee:2019wcg, Goyal:2020bkm, Finke:2021znb}, localization gravitational-wave sources~\citep{hannuksela2020localizing, Yu:2020agu}, searches for intermediate-mass black holes~\cite{Lai:2018rto}, primordial black holes and low mass dark matter halos~\citep{Diego:2019rzc, Oguri:2020ldf}, the mass sheet degeneracy problem~\citep{Cremonese:2021puh}, and precision cosmology~\citep{Sereno:2011ty,Liao:2017ioi,Liao:2019hfl,Cao:2019kgn,Li:2019rns,Hou:2019dcm,Mukherjee:2019wfw}. 

In the strong lensing regime, when a GW is lensed by a galaxy or a galaxy cluster, the geometrical optics approximation can be used to solve for the trajectories, magnifications, arrival times, and types of the lensed images \citep{Takahashi:2003ix,Dai:2016igl, Ng:2017yiu, Li:2018prc, Oguri:2018muv,Smith:2017mqu,Smith:2018gle,Smith:2019dis,Robertson:2020mfh,Ryczanowski:2020mlt}. 
The gravitational-wave images would differ in arrival times, amplitudes and overall phases~\citep{Wang:1996as, Haris:2018vmn,Ezquiaga:2020gdt,Dai:2020tpj,Liu:2020par,Wang:2021,Lo:2021nae,Janquart:2021nus}. 
The frequency evolution of the GW is still the same for these macroimages. 
Therefore, searches based on matched filtering \citep{Li:2019osa, McIsaac:2019use} and Bayesian analysis \citep{Haris:2018vmn, Hannuksela:2019kle, Dai:2020tpj, Liu:2020par, Lo:2021nae, Janquart:2021nus}.  are used in the search of GW strong lensing 

However, when GWs in the frequency band of LIGO are lensed by a stellar mass object, so-called wave optics effects become significant~\citep{Deguchi:1986zz, Nakamura:1997sw, Takahashi:2003ix}. 
In particular, as the object's Schwarzschild radius becomes comparable to the wavelength of the GW, diffraction and interference effects will become important~\citep[e.g.][]{nakamura1999wave,Takahashi:2003ix,Diego:2019lcd,Cheung_2021,mishra}. 
Indeed, recent studies have shown how wave optics effects can alter microlensing signatures when gravitational waves are lensed by stellar mass lenses~\citep{Christian:2018vsi, Diego:2019lcd, mishra, Cheung_2021,Meena:2022unp}.

There has been progress in understanding the detection of microlensing, and recent literature points towards strong lensing making microlensing detections easier. 
When microlensing takes place for a strongly lensed image, microlensing signatures in the GWs become more prominent~\citep{Diego:2019lcd,Diego:2019rzc,Cheung_2021,mishra,seo2022improving,Meena:2022unp}. 
In addition to the increase in the microlensing detection probability and magnitude provided by strong lensing, \citet{seo2022improving} demonstrated how strong lensing could be used in parameter estimation to further improve our ability to detect microlensing.
Therefore, understanding the interplay between strong lensing and microlensing will likely play a key role in realistic detections.

\begin{figure}
    \centering
    \includegraphics[width=8cm]{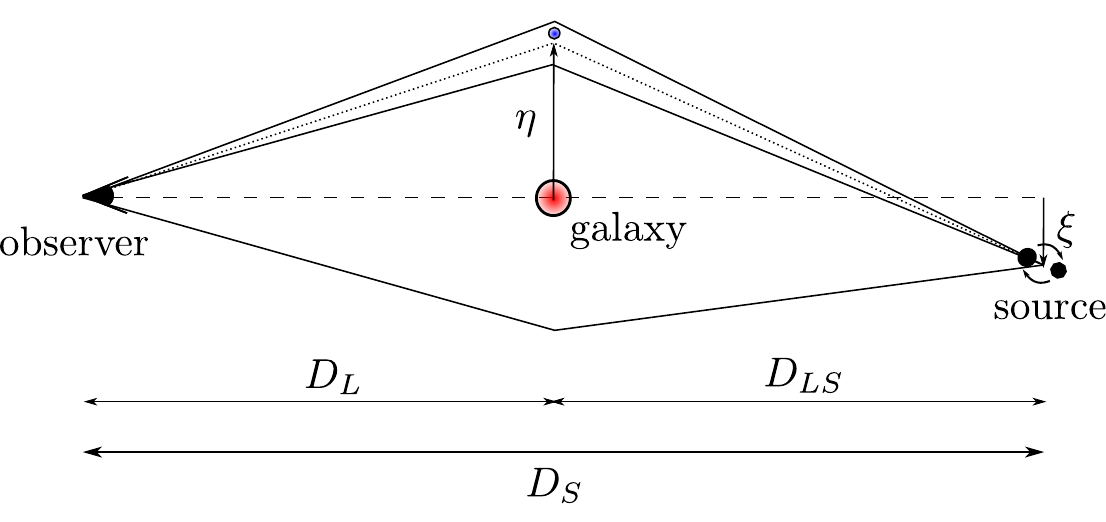}
    \centering
    \includegraphics[height=5cm,width=7cm]{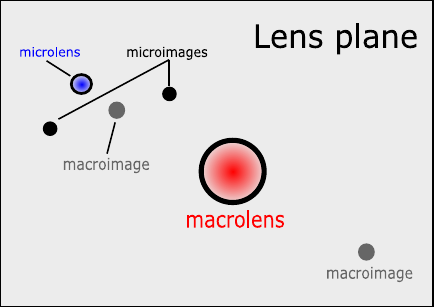}
    \caption{Schematic diagram of the compound lens model.
    (Top panel): In the geometrical-optics limits, GWs are treated as rays propagating from the source to the observer. 
    $D_S$, $D_L$ are the distances from the observer to the source and the galaxy-scale lens. 
    $D_{\rm LS}$ is the distance between the source and the macrolens.
    (Bottom panel): In this example, the macrolens (red-bounded circle, which is a galactic halo in the cases studied in this work) lenses the signal into two macroimages (grey circles).
    One macroimage is formed near a microlens (blue-bounded circle) and is further lensed into multiple microimages (black circles).
    The distance between the macroimage and the microlens is called the micro-impact parameter $\vec{\eta}_{\rm m}$.}
    \label{fig:schematic}
\end{figure}

In prior work,~\citet{Diego:2019lcd} studied the effects of macromodel lensing on microlensing, primarily in the limit of extreme magnification. \citet{Cheung_2021} studied how type-\Rom{1} images produced by the macromodel are altered by stellar-mass lenses in the regime of low magnification, concluding that the GWs will likely propagate through such small-scale lenses unimpeded. 
\citet{mishra} studied the effects of macromodel on fields of microlenses in various configurations and stellar densities.
Here, focusing on the low magnification regime (macromagnification $\mu  \lesssim 10$), we investigate type-\Rom{2} images in the context of the microlensing detection and science case and carry out parameter estimation for the first time. 
We find that type-\Rom{2} images produce larger microlensing-induced mismatches (the mismatch between a strongly lensed image and the microlensed version of the same image) than type-\Rom{1} images, supporting prior findings by~\citet{mishra}, and that for the case of microlensing of type-\Rom{1} images, the microlens parameters will generally not be possible to recover with the current parameter estimation searches.
\footnote{In \cite{seo2022improving}, the authors showed that the microlens mass for type I images could be recovered if the injected waveform is lensed by an isolated point mass with an additional constant magnification.
This is different from the cases considered in this work, where the injected microlensed waveforms are computed considering the full compound lens model as a whole, for which the microlens parameters cannot be recovered for the type-\Rom{1} image case.}

This paper is organized as follows. 
In Section \hyperref[sec:methods]{2}, we review the geometrical optics and wave optics calculations relevant to the compounded lens model comprising of a galaxy-scale lens and a stellar-mass lens. 
In Section \hyperref[sec:results]{3}, we then show the wave optics amplification factor around a type-\Rom{2} image. 
We also present the deviation induced in the waveform due to the microlens and discuss the detectability of such deviations. 
We discuss the recovery of the microlens's parameter and the detectability of microlensing of signals from the compounded lens model with Bayesian analysis.
In Section \hyperref[sec:implication]{4}, we will discuss implications of our work on the detection of microlensing.
Throughout the paper, we will use geometrical units with $c = G = 1$.

\section{Methods}
\label{sec:methods}

A schematic diagram of the lens system is shown in Fig. \ref{fig:schematic}. 
A GW source is positioned at an angular diameter distance of $D_S$. 
$D_L$ is the angular diameter distance from the observer to the lens. 
Since $D_S$ and $D_L$ are very large compared to the scale of the lens, the thin-lens approximation is applied and we assume that the lensing of GWs occurs in a single plane, which we call the lens plane. 
In a strong lensing scenario, when a GW in the LIGO frequency band is lensed by a sufficiently large lens, for example a galaxy, it could produce multiple images with variable magnifications, arrival times, and Morse phases. 
We call this phenomenon `macrolensing.'
The galaxy-scale lens is the `macromodel,' and the separate images are the `macroimages.'
Such nomenclature is used to contrast the lensing effects contributed by the macromodel from that of the `micromodel':
if one of the macroimages lies close to a stellar-mass object, for example a star or a stellar-mass black hole, that macroimage could be further split into additional images. 
We call this `microlensing,' the stellar mass lens the `micromodel,' and the new images the `microimages.'

By considering `images' of lensed GWs, we have implicitly invoked the geometrical optics approximation, treating GWs as rays passing only through the image locations. 
In fact, as we will see below, when the wavelengths of GWs observed by LIGO are comparable with the characteristic length-scale of the microlens, the geometrical optics approximation no longer holds, and we have to take into account diffraction effects. 
Then, we will no longer consider the GWs as rays that only pass through the image positions but passing through every point on the lens plane before arriving at the observer and interfering with itself to produce the GW signal observed. 
Nevertheless, below we will still define the image positions to be the stationary points of the time-delay function. 
Due to the stationary-phase approximation, GWs that pass through points nearer to an image contribute more to the final observed GW signal. 

\FloatBarrier
\begin{figure*}
    \centering
    \includegraphics[height=8.5cm, width=18cm]{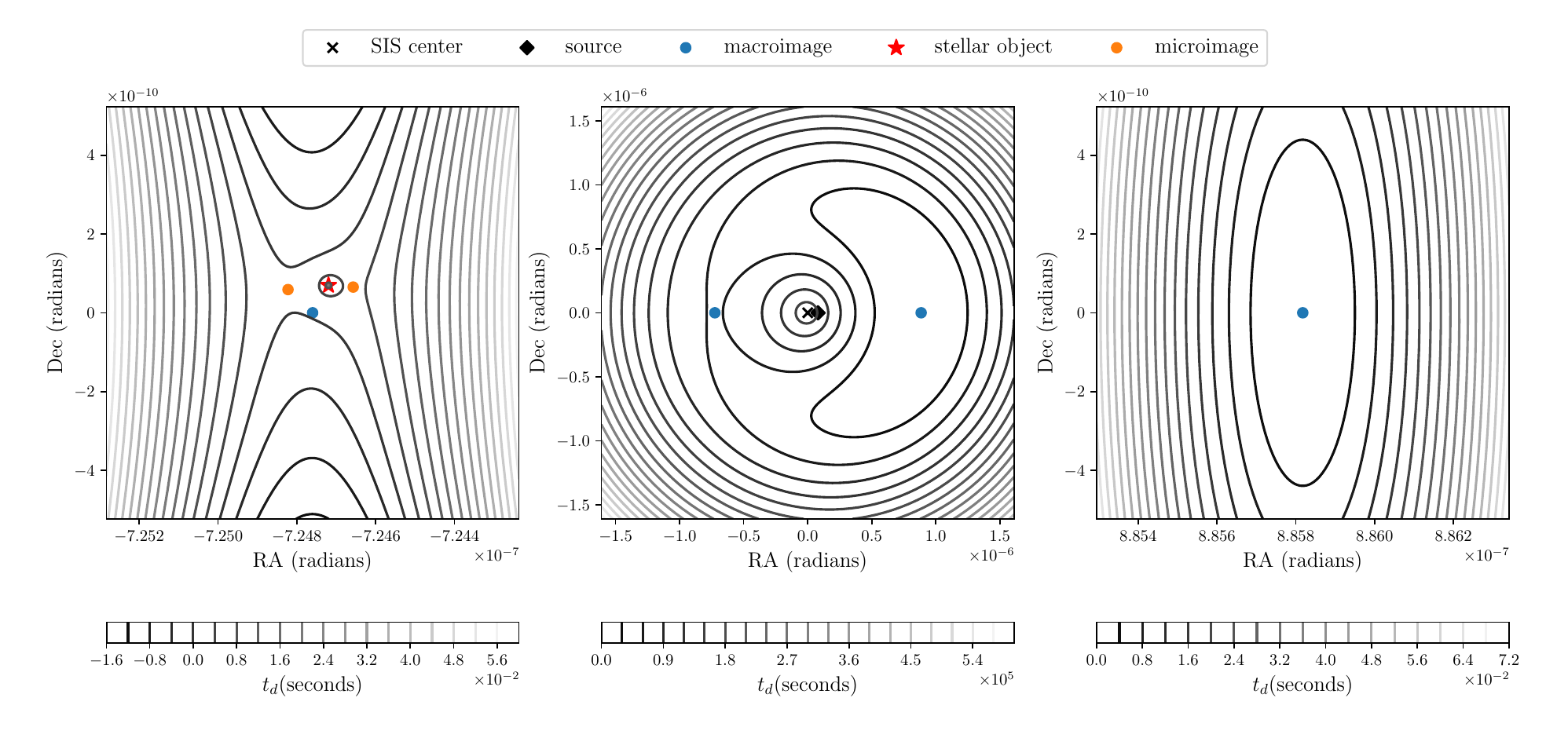}
    \caption{(Centre panel): The lens plane of an SIS lens macromodel of mass = $10^{10}{\rm M}_{\odot}$. The source position is projected onto the lens plane and located at a macro-impact parameter of $\vec{\eta}_{\rm SIS}$ away from the centre of the SIS lens.
    The macromodel produces a saddle point image (blue dot on the left) and a minimum image (blue dot on the right).
    (Left-hand panel): Zoom-in region around the saddle point image.
    The macroimage is split into two microimages in the presence of a microlens (red star).
    At the saddle point macroimage, the time contours do not form a closed loop locally.
    If the arrival time of the saddle point image is marked as 0, the left-right hyperbolic contours resemble a positive time delay relative to the saddle point, while the up-down ones represent a negative time delay. 
    The density of left-right hyperbolas is higher than that of up-down hyperbolas.
    (Right-hand panel): Zoom-in picture of a type-\Rom{1} macroimage.
    Unlike type-\Rom{2} macroimage, the time contours around type-\Rom{1} are closed locally.
    The time contours are measured in seconds.
    For the left-hand and right-hand panel, $t_{\rm d} = 0$ is chosen by convention} to be the original arrival time of the macroimage without microlensing.
    \label{fig:contours}
\end{figure*}

\subsection{Diffraction Integral for amplification factor}

To calculate diffraction effects, we treat GWs as rays. 
When the rays pass through the lens plane, they are bent by the gravitational field of the lens. 
Accounting for the geometry of the path and also the time delay due to the gravitational field, different rays arrive at the observer at different times. 
The arrival time of an image can be calculated from a time delay function $t_{\rm d}(\vec{\xi}, \vec{\eta})$, parameterized by source position $\vec{\xi}$ on the source plane and the image position $\vec{\eta}$ on the image plane.  
For convenience, the analysis of $\vec{\xi}$ and $\vec{\eta}$ will be worked in units normalized by the Einstein radius 
\begin{equation}
    \theta_{\rm E} = \sqrt{\frac{4M_{\rm L} D_{\rm L} D_{\rm LS}}{D_{\rm S}}}\, ,
\end{equation}
where $M_L$ is the total mass of the lens and $D_{\rm LS}$ is the angular distance between the lens and the source.
The Einstein radius is the characteristic length-scale of the lens. 
In normalized units, $\vec{x} = \vec{\xi}/\theta_{\rm E}$, $\vec{y} = \vec{\eta}/\theta_{\rm E}$, the cosmological time delay function is given by 
\begin{equation}
    t_{\rm d}(\vec{\xi}, \vec{\eta}) = \frac{D_{\rm S} \theta_{\rm E}^2}{D_{\rm L} D_{\rm LS}}(1+z_{\rm L}) T_{\rm d}(\vec{x}, \vec{y})\,,
\end{equation}
with the time delay function $T_{\rm d}$ given by
\begin{equation}
    \begin{split}
        T_{\rm d}(\vec{x}, \vec{y}) &= T_{\rm geom} + T_{\rm g} \\
        &= \frac{1}{2} \lvert \vec{x}-\vec{y} \rvert^2 - \psi(\vec{x}) - \phi_{\rm m}(\vec{y})\,,
    \end{split}
\end{equation}
 $\psi(\vec{x})$ is the deflection potential and $\phi_{\rm m}(\vec{y})$ is the arrival time of the first image.
The factor $(1+z_{\rm L})$ is included to consider the cosmological distance, where $z_{\rm L}$ is the lens' redshift.

The lensed waveform is a product of an amplification factor and the unlensed waveform: 
\begin{equation} \label{eq:amplification}
    h_{\rm L}(f) = F(f)h_{\rm UL}(f)\,,
\end{equation}
with the (complex) amplification factor 
\begin{equation}\label{eq:waveopt}
    F(f) = \frac{D_{\rm S} \theta_{\rm E}^2 (1+z_{\rm L})}{D_{\rm L} D_{\rm LS}} \frac{f}{i} \int {\rm d}^2\Vec{x} e^{2\pi f t_{\rm d}(\Vec{x},\Vec{y})}\,.
\end{equation}
In terms of the dimensionless frequency $\rm w = 8\pi M_L (1+z_L)f$, the amplification factor becomes 
\begin{equation}\label{eq:diffint}
    F(w) = -iw \int {\rm d}^2\Vec{x} e^{iwT_{\rm d}(\Vec{x},\Vec{y})}\,.
\end{equation}
Finally, taking the inverse Fourier transform of the amplification factor, 
\begin{equation}
    F(t) = \int {\rm dxdy} \delta(T_{\rm d}(\Vec{x},\Vec{y})-t)\,.
\end{equation}
The diffraction integral is evaluated on the $x - y$ lens plane. 
To solve the diffraction integral, we cover the lens plane with a dense grid.  
For each grid point, we evaluate the time delay function at that point. 
The histogram of these time delay values with respect to $t$ would then be $F(t)$, and the inverse Fourier transform of its derivative gives $F(f)$~\citep{Diego:2019lcd}. 

\subsection{Geometrical optics approximation on microlensing}

For $w \gg 1$, one can adopt the geometrical optics approximation. 
By using Fermat's principle, the images produced by the lens are stationary points of the time delay function. 
The diffraction integral reduces to a sum over images on the lens plane~\citep{schneider1992gravitational}
\begin{equation}\label{eq:geomapx}
    F(f) = \sum_j \lvert \mu_{\rm j} \rvert^{1/2} e^{2\pi if t_{\rm d,j} - i \pi n_{\rm j}}
\end{equation}
where $j$ denotes the $j$th image, $\mu_{\rm j}$ is the magnification of the $j$th image and $n_{\rm j}$ is the Morse index of the $j-$th image. $n_{\rm j}$ = $0$, $1/2$, $1$ for a minimum, saddle point and maximum point of $T_{\rm d}$, respectively. 
We solve the positions, magnifications and Morse indices of images with the \textsc{LensingGW} package \citep{pagano2020lensinggw}.

The geometrical optics limit is sufficient for strong lensing of GWs in the LIGO frequency band by a galaxy-scale lens. 
In this limit, the size of the lens is much larger than the wavelength of the GW.
In microlensing, the lens size (Schwarzschild radius) can be comparable to the wavelength of GW. 
Thus, it introduces diffraction effects, which we can account for by examining the whole diffraction integral on the lens plane without making use of the geometrical optics approximation. 

Note that the microimages are defined to be located at the stationary points of the time delay function. 
Therefore, the positions, magnifications and morse indices of the microimages can still be obtained. 
Using these microimages, we can investigate microlensing in the absence of full wave optics effects, allowing us to quantify the wave optics effects as the difference. %

\subsection{Integration window size}
The diffraction integral equation. (\ref{eq:diffint}) is integrated by estimating the area between two contour lines with consecutive values of time delay. 
In theory, this would require us to use an area window in the lens plane that encloses all of the contour lines with a time delay close to that of the arrival time of the macroimage under consideration. 
The contour lines next to type-\Rom{2} images would extend far away from the image itself because it is a saddle point. 
In fact, if we use an area window that encloses all of these contour lines, it will need to span the entire macromodel itself (centre panel of Fig.~\ref{fig:contours}).
It can be seen in Fig. \ref{fig:contours} that the time delay contours do not close immediately around the saddle point, extending out of the window of integration instead. 
When computing microlensing effects, we require a very fine resolution in time delay, so a large window for integration would be impractical. 
Therefore, as shown in the left-hand panel of Fig.~\ref{fig:contours}, we use a small integration window that encloses only the immediate vicinity of the image. 
Due to the stationary phase approximation, we know that the region near the image would contribute the most to the integral, so a small window around the image should give us a good approximation. 

To compute the diffraction integral (\ref{eq:diffint}) for type-\Rom{2} macroimages to within acceptable margin or error (e.g. 1 per cent), a large integration domain has to be used, which in turn increases the computational time. 
The area of the integration domain for type-\Rom{2} is almost three orders of magnitude (number of pixels $N$ = 3.0 $\times 10^5$ at each side) larger than that of type-\Rom{1} ($N$ = 1.0 $\times 10^4$) in order to account for the sufficient contribution of time delay function to converge the amplification factor, leading to a much higher computational cost. 
Beyond the larger computational domain, the recipe for computing type-\Rom{2} image amplification factors can be done similarly to the type-\Rom{1} case.

\begin{figure}
    \centering
    \includegraphics[height=6cm, width=7.5cm]{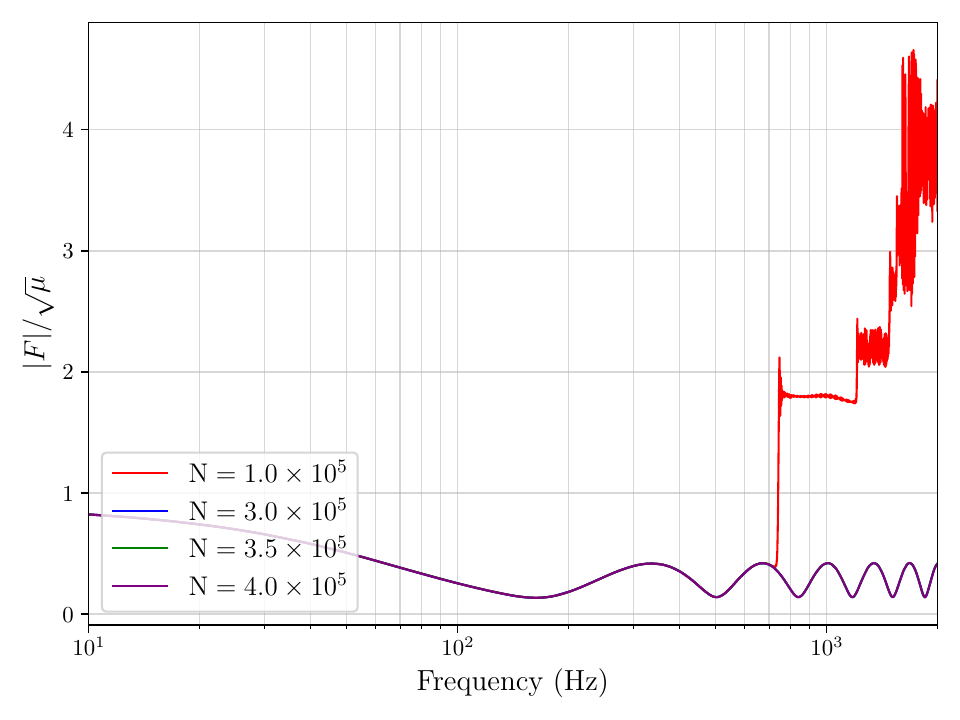}
    \includegraphics[height=6cm, width=8cm]{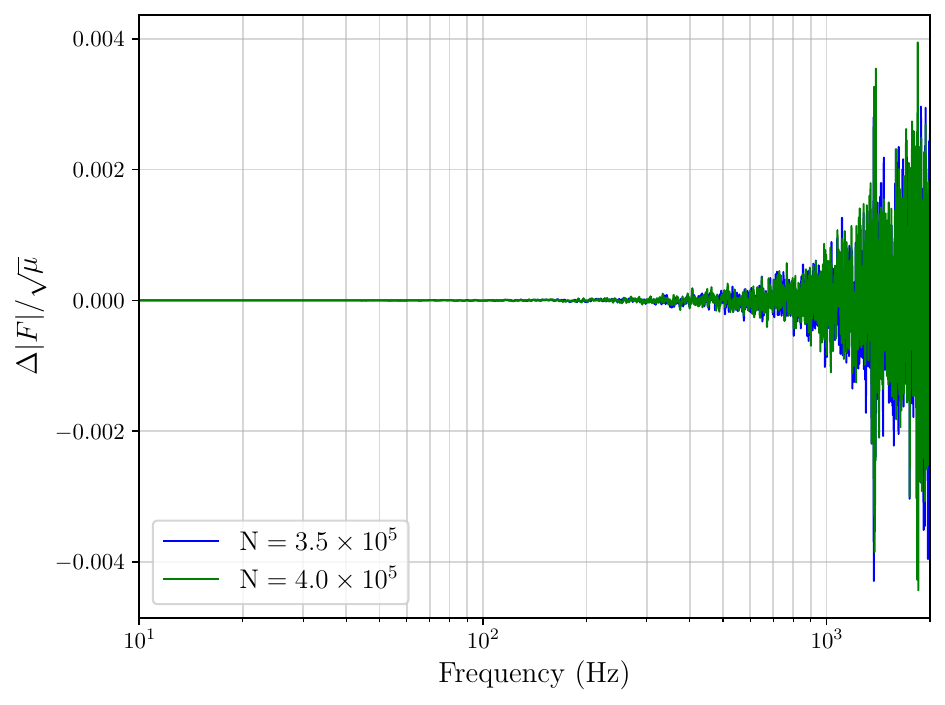}
    \caption{(Top panel): The amplification factors with the number of pixels at the sides being $N$ = (1.0, 3.0, 3.5, 4.0) $\times 10^5$. $F(f)$ diverges at 1.0 $\times 10^5$ and converges beyond 3.0 $\times 10^5$. A large number of pixels with at least  3.0 $\times 10^5$ is required for microlensing analysis around a type-\Rom{2} macroimage. (Bottom panel): The deviation of amplification factors of $N$ = (3.5, 4.0) $\times 10^5$ with respect to $N$ = 3.0 $\times 10^5$ used for the analyses. The subpercent differences suggests the nonnecessity of further increasing the integration window. We adopt the number of pixels as 3.0 $\times 10^5$ for the balance between computational efficiency and accuracy of the result.}
    \label{fig:difference}
\end{figure}

We justify the choice of the integration area by extending the window size for the integration. 
If our integration approximates the solution well enough that increasing the grid size does not alter the final result, further extension of integration window would not be required. 
On the other hand, if the chosen integration area is too small, window extension would give a significantly different, and more accurate, amplification curve. 
Checking such convergence allows us to determine the validity of the chosen integration area. 
As shown in Fig. \ref{fig:difference}, the amplification factor with our window choice deviates below 1 per cent when the window size increases by a third. 
That is, the error due to using a restricted window is acceptable for our purposes.

However, if we would like to obtain an amplification curve accurate to the sub-percent level, then we would have to integrate over a much larger area in the lens plane.
This would be the case if we want to measure the parameters of a microlens with future GW detectors of higher sensitivity, where a deviation in 1 per cent of the waveform will affect parameter estimation.
Alternative methodologies to evaluate the diffraction integral have to be employed for these use-cases.

\section{Results}
\label{sec:results}
In this section, we quantify the effects of stellar microlenses on GWs lensed by a galaxy, specifically for the type-\Rom{2} macroimage. 
The GW source is placed at a redshift of $z_{\rm s} = 1$ and the macrolens is placed at $z_{\rm l} = 0.5$. 
For the macromodel, we adopt a singular isothermal sphere (SIS) with a lens mass $10^{10} {\rm M}_{\odot}$ with an Einstein radius of $\theta_{\rm SIS}$. 
Such a macromodel will produce multiple GW images if the source is placed within $< \theta_{\rm SIS}$ away from the centre of the lens galaxy.

On the lens plane, the source is placed at an angular distance of $\eta_{\rm SIS} = 0.1 \theta_{\rm SIS}$ away from the centre of the lens galaxy (as measured after projecting the source on the lens plane), which produces a macromagnificaton of $\mu = 9$ for the type-\Rom{2} macroimage.
Since the source is within an Einstein radius from the line of sight to the galaxy, there will be two macroimages, 
one of type-\Rom{1} and the other of type-\Rom{2}. 
Then, we place a point-mass microlens at an angular distance $\vec{\eta}_{\rm m}$ away from the macroimage position. 
We call $\vec{\eta}_{\rm m},$ the `micro-impact parameter', as it is the impact parameter of the type-\Rom{2} macroimage (if it were a single ray) with respect to the microlens.
Instead of working with $\vec{x}$ and $\vec{y}$, we work with the their unnormalized counterparts $\vec{\xi}$ and $\vec{\eta}$ for convenience.

\subsection{Wave optics approximation}

Fig. \ref{fig:contours} shows the set-up of the micromodel and the macromodel, where we have macro-impact parameter $\vec{\eta}_{\rm SIS} = 0.1 \theta_{\rm SIS}$, micro-impact parameter $\vec{\eta}_{\rm m} = 1 \theta_{\rm m}$ and microlens mass $m = 100 {\rm M}_{\odot}$.
The microlens is placed at $60 \degree$ counter-clockwise above the line connecting the macroimage and the SIS lens, shown as a star in the left-hand panel. 
The original type-\Rom{2} macroimage, shown as a blue dot in the left-hand panel, is split into two type-\Rom{2} microimages, shown as orange dots.
If we did not use a singular microlens like the point mass lens we used here, the singular point of the time delay function will be replaced by a maximum image, and we would see three microimages.

With the \textsc{LensingGW} package, the positions, magnifications and morse indices are found for each microimage. 
They are used for computing the amplification factor in the geometrical optics approximation \eqref{eq:geomapx}. 
On the other hand, the full wave optics analysis of the system is calculated with \eqref{eq:waveopt}.

\begin{figure}
    \centering
    \includegraphics[height=6cm, width=8cm]{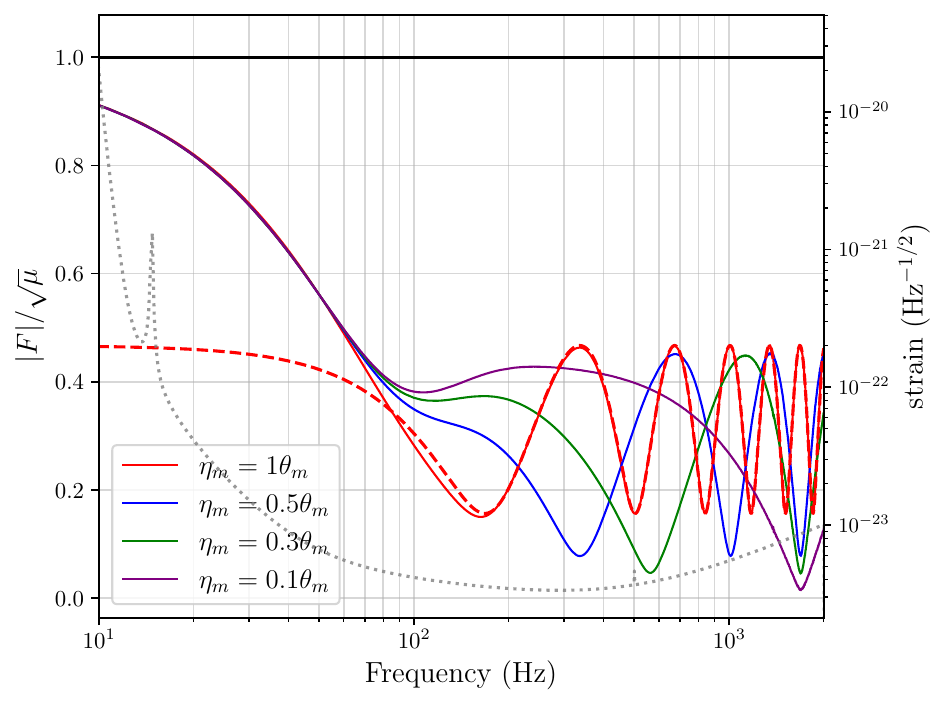}
    \centering
    \includegraphics[height=6cm, width=8cm]{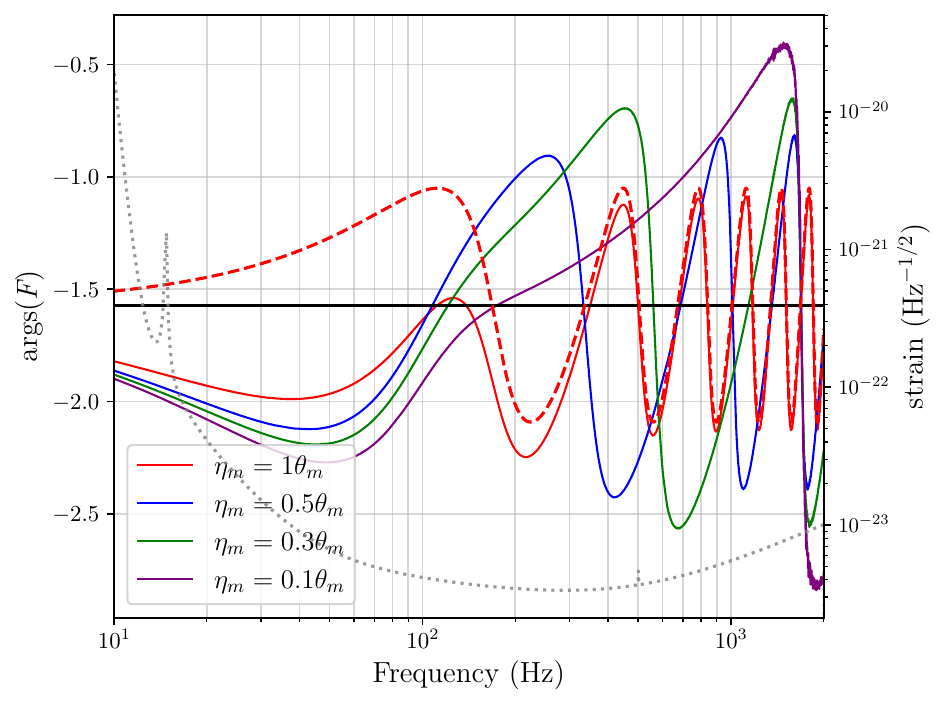}
    \caption{
    The normalized amplitude and the phase of the amplification factor with the micro impact parameter $\eta_{\rm m} \in \{0.1, 0.3, 0.5, 1\} \theta_{\rm m} $ and a fixed microlens lens mass $m$ = 100 ${\rm M}_{\odot}$. 
    The microlens is placed at $60 \degree$ from the polar axis.
    Wave optics suppresses the amplification factor the macromagnification and the morse phase of $\frac{\pi}{2}$ of the type-\Rom{2} macroimage.
    The red dashed line shows the results obtained by the geometrical optics approximation for the case $\eta_m = 0.1 \theta_m$.
    For readability, the geometrical optics curves are not shown for the other cases, but they show a similar behavior, where the wave optics curves are always suppressed compared to them.}
    \label{fig:variation_distance}
\end{figure}

\begin{figure}
    \centering
    \includegraphics[height=6cm, width=8cm]{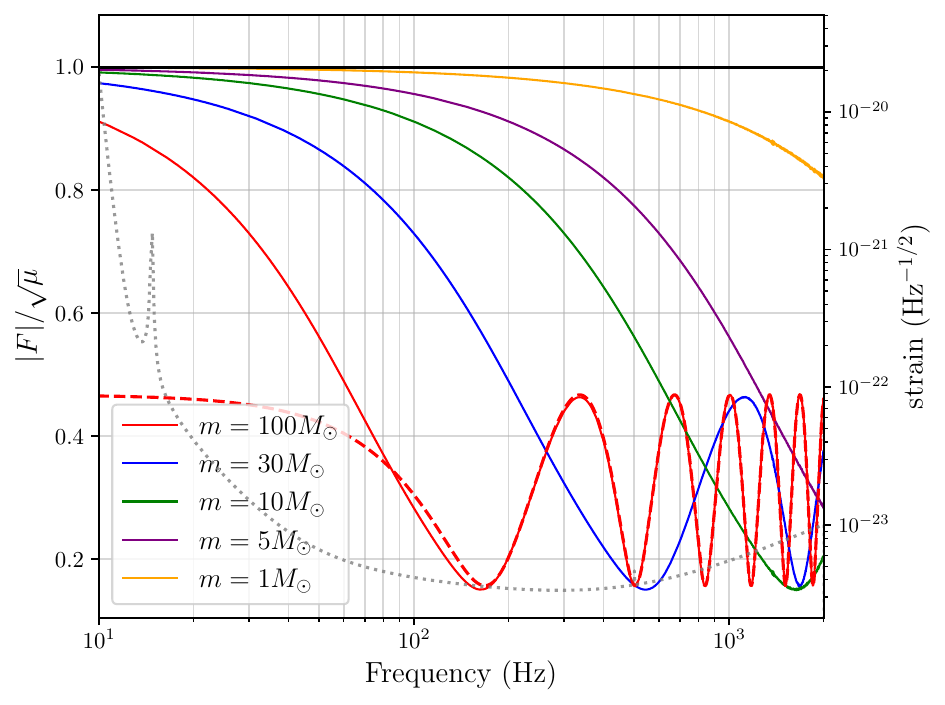}
    \centering
    \includegraphics[height=6cm, width=8cm]{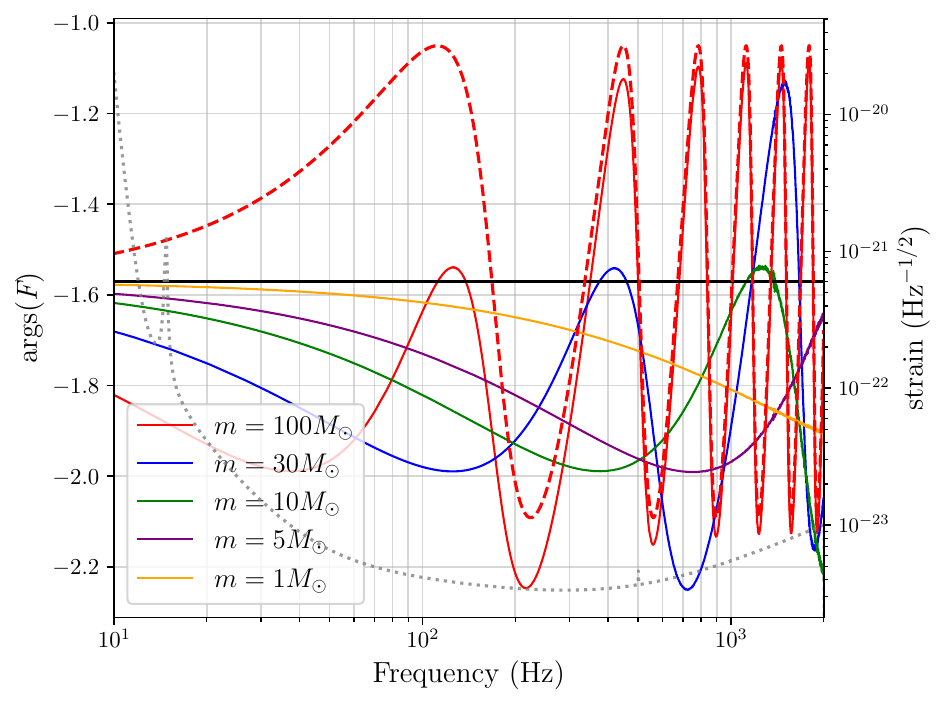}
    \caption{Same as Fig. \ref{fig:variation_distance}, but with microlens mass $m \in \{1, 5, 10, 30, 100\} {\rm M}_{\odot}$ with $\eta_{\rm m}$ = 1 $\theta_{\rm m}$ away from the microlens.
    Amplification factors are suppressed at the low-frequency regime.
    The red dashed line shows the results obtained by the geometrical optics approximation for the case $m = 100 M_\odot$.}
    \label{fig:variation_mass}
\end{figure}

Figs \ref{fig:variation_distance} and \ref{fig:variation_mass} show the magnification factors $F(f)$ using the full wave optics calculations and the geometrical optics approximation.
The amplification factor is normalized by $\sqrt{\mu}$ where $\mu$ is the magnification of the type-\Rom{2} macroimage without microlensing effects, such that in the absence of a microlens the amplification factor reaches unity. 
The geometrical optics approximation deviates from the true amplification with wave optics in the low frequency regime. 
Indeed, it is vital to consider the macromodel in conjunction with microlensing in most strong lensing set-ups \citet{Diego:2019lcd,mishra}. 
For illustrative purposes, we have shown the PSD of aLIGO \citep{PhysRevD.93.112004}.
The wave optics suppression happens at the low frequency region which is the sensitive region of the detectors.

Similarly to \citet{mishra}, we observe that the normalized amplification factor $\left| F\right|/\sqrt{\mu}$ in full wave optics treatment reduces to unity when $f \to 0$. 
The same convergence occurs for the phase $\arg(F)$, which reduces to a phase shift of $\pi/2$ when $f \to 0$, which is the expected result for a type-\Rom{2} macroimage. 
Low frequency corresponds to GWs of wavelength larger than the Schwarzschild radius, while diffraction occurs when its wavelength is comparable to the Schwarzschild radius. 
Thus, GWs with longer wavelengths are more suppressed by the diffraction effects.
Such a phenomenon is similar to that of type-\Rom{1} image microlensing \citep{Cheung_2021}, and in line with previous findings~\citep{mishra}. 

We then investigate how the location and the mass of the microlens affects the wave optics suppression. 
Figs \ref{fig:variation_distance} and \ref{fig:variation_mass} show the amplification factor in different scenarios with variable location and mass of the microlens. 
Wave optics suppression presents regardless of the micro-impact parameter and the mass of the microlens, with $F$ converging to $1$ at low frequencies.
Interestingly, the low frequency behavior is asymptotically independent of changes in the micro-impact parameter $\eta_{\rm m}$, which is an effect similar to the type-\Rom{1} macroimage case shown in Fig. 7 of \cite{Cheung_2021}.

\subsection{Demagnification for the microlensed type-\Rom{2} image}

We examine the effective magnification due to microlensing around type-\Rom{2} macroimages. 
We define the effective micro magnification as
\begin{equation}
    \mu_{\rm m} = \frac{\langle h_{\rm L, full}, h_{\rm L, full} \rangle}{\langle h_{\rm L, macro}, h_{\rm L, macro} \rangle} \,,
\end{equation}
where 
\begin{equation}
    \langle a, b \rangle = 4 \, {\rm Re}{\int_{f_{\rm low}}^{f_{\rm high}} \frac{a(f)b^{*}(f)}{S_{\rm n}(f)}} {\rm d}f\,.
\end{equation}
is an inner product weighted by the power spectral density (PSD) $S_{\rm n}(f)$ of LIGO \citep{PhysRevD.93.112004}. 
The dependency of frequency of amplification factor is removed by taking the noise weighted inner product. 
$\mu_{\rm m}$ can be seen as the microlens's contribution to the magnification of the lensed GW.
As the magnification is vital in the localization and follow-up cosmography of strongly lensed GWs \citep[e.g.][]{Sereno:2011ty,hannuksela2020localizing,Yu:2020agu}, if $\mu_{\rm m}$ significantly deviates from $1$, microlensing might contaminate the signal to a degree that would affect these studies. 

\begin{figure}
    \centering
    \includegraphics[height=6cm, width=9cm]{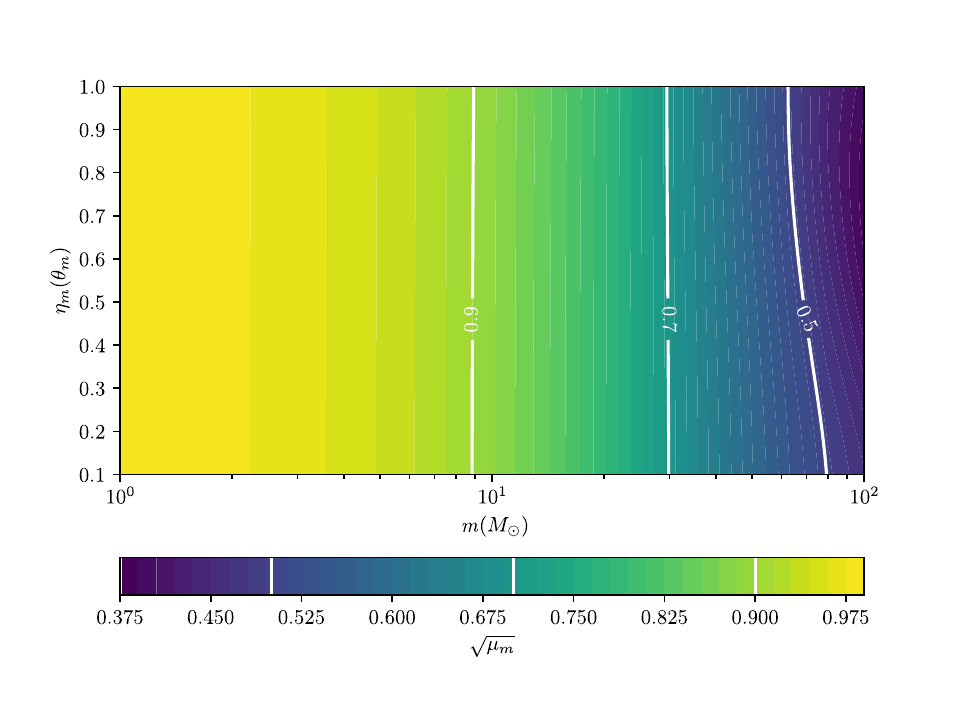}
    \caption{Contour plot of the effective micro-amplification $\sqrt{\mu_{\rm m}}$ as a function of the micro impact parameter $\eta_{\rm m}$ and the microlens mass $m$. $\sqrt{\mu_{\rm m}}$ is approximately independent of $\eta_{\rm m}$ for $m \lesssim 80 {\rm M}_{\odot}$.}
    \label{fig:effectivemu}
\end{figure}

Fig. \ref{fig:effectivemu} gives the effective magnification as a function of the micro impact parameter  $\eta_{\rm m} \in [0.1, 1.0] \theta_{\rm m} $ and mass of microlens $m \in [0,100] {\rm M}_{\odot}$.
Note that the effective magnification is normalized by the macroimage magnification. 
When $\sqrt{\mu_{\rm m}} < 1$, the microlens demagnifies the image. 
We used the \textsc{IMRPHenomXPHM} waveform generated by $(30, 30)$ ${\rm M}_{\odot}$ of a binary  black hole merger with \textsc{pycbc} \citep{pycbc}. 
As the microlens mass becomes small ($m \lesssim 10 {\rm M}_{\odot}$), $\mu_{\rm m}$ reaches unity. 
This means that the effect of microlensing is insignificant for small $m$; the wave travels past the microlens unimpeded. 
$\sqrt{\mu_{\rm m}}$ stays smaller than 1 by  $\lesssim$ 10 per cent for $m \lesssim 10 {\rm M}_{\odot}$. 
In general, the microlensing for a type-\Rom{2} image would give us a demagnified waveform compared to the clean macrolensed case.

Let us then investigate how $\eta_{\rm m}$ and $m$ alter the waveform and its detectability using mismatch analysis. 
When the mismatch in the waveform due to the microlens is large (often 3 per cent is used as the cut-off for typical LIGO events), the microlensing effects would be distinguishable. 
The deviation between the lensed and unlensed waveforms is defined by the waveform overlap which is given by the normalized inner product between the two waveforms
\begin{equation}
    \mathcal{O}[h_{\rm L, full}, h_{\rm L, macro}] = \frac{\langle h_{\rm L, full}, h_{\rm L, macro} \rangle}{\sqrt{\langle h_{\rm L, full}, h_{\rm L, full} \rangle \langle h_{\rm L, macro}, h_{\rm L, macro} \rangle}},
\end{equation}
where $h_{\rm L, full}$ is the full microlensed waveform, while $h_{\rm L, macro}$ is the waveform that is only macrolensed but not microlensed.
We then define the match $\mathcal{M}[h_{\rm L, full}, h_{\rm L, macro}]$  to be the maximum of $\mathcal{O}[h_{\rm L, full}, h_{\rm L, macro}]$ over the starting time and the overall phase factor of the signal, and the mismatch to be 1-$\mathcal{M}[h_{\rm L, full}, h_{\rm L, macro}]$ which quantifies the frequency dependent deviation induced in the waveform by microlensing. 
The match is computed using \textsc{pycbc}. 

\begin{figure}
    \centering
    \includegraphics[width=0.49\textwidth]{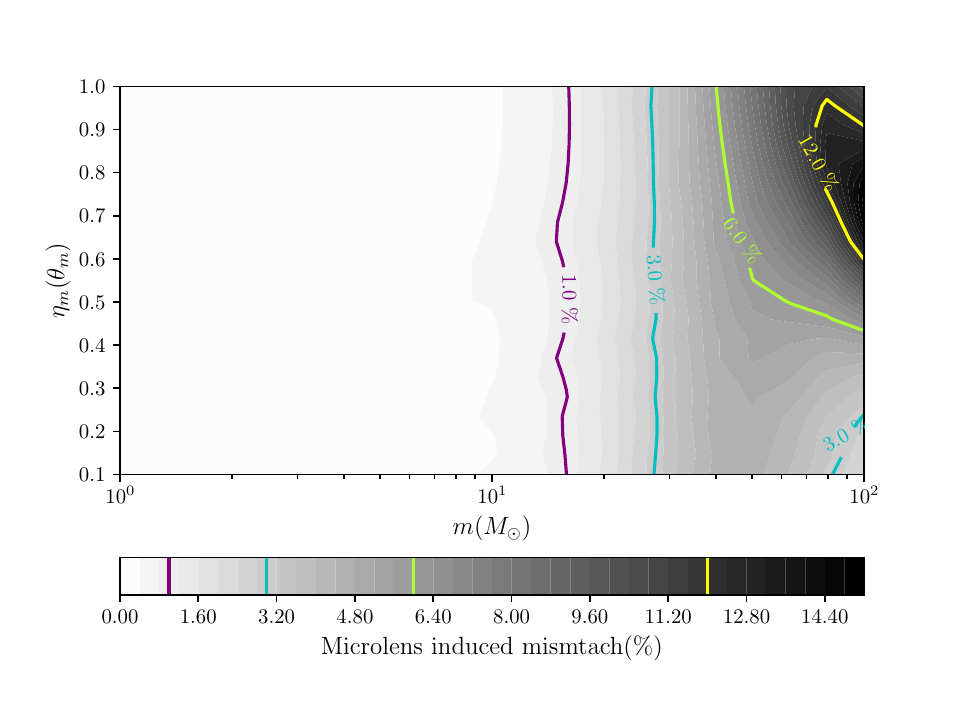}
    \caption{The mismatch of type-\Rom{2} $h_{\rm L, full}$ and $h_{\rm L, macro}$ in the parameter space of $\eta_{\rm m}$ and $m$. The contour lines of mismatch $\in \{1.0, 3.0, 6.0, 12.0\}$ pre cent are shown. The mismatch is subpercent for $m \lesssim 10{\rm M}_{\odot}$.}
    \label{fig:mismatch}
\end{figure}

Fig. \ref{fig:mismatch} shows the contour plot of mismatch between waveform with and without microlensing as a function of the microlens mass $m$ and the micro impact parameter $\vec{\eta}_{\rm m}$. 
The mismatch is of sub-percent level for m $\lesssim 10 {\rm M}_{\odot}$. 
As for $m \gtrsim 20 {\rm M}_{\odot}$, the mismatch is greater than 3 percents. 
This suggests that the microlensing around a type-\Rom{2} macroimage would introduce a potentially non-negligible effect to the measurement of the GW parameters for high mass microlenses, as hinted by~\citet{mishra}. 
\footnote{
This paper uses a different physical set-up from \citet{mishra}.
We place a single microlens near the macroimage and vary the microlens's mass around 100 ${\rm M}_{\odot}$, while
\citet{mishra} computed the mismatch of injecting field of microlens following Chabrier IMF function, which concentrates in low mass compact objects lower than $10 {\rm M}_{\odot}$, in magnifications as high as $\gtrsim 100$.
Hence, our results are expected to be quantitatively different from \citet{mishra}, e.g. in the microlens-induced mismatch as in Fig.~\ref{fig:mismatch}.
}
As opposed to microlensing of the type-\Rom{2} image, the mismatch due to microlensing on the type-\Rom{1} image is negligible in the measurement as it introduces mismatch $\lesssim 1 \%$ for the same parameter space \citep{Cheung_2021}.  
In the Appendix, we show that the mismatch varies with the relative angle between the galaxy and the microlens due to the saddle point geometry of the type-\Rom{2} image.
Regardless of the relative angle, the mismatches for type-\Rom{2} images are greater than that of type-\Rom{1} images.
Therefore, microlensing by type-\Rom{2} images may be easier to detect than a type-\Rom{1} image with similar signal to noise ratio (SNR).


\subsection{Parameter estimation on the wave optics waveform}

\begin{figure*}
    \begin{multicols}{2}
        \centering
        \includegraphics[height=6cm, width=9cm]{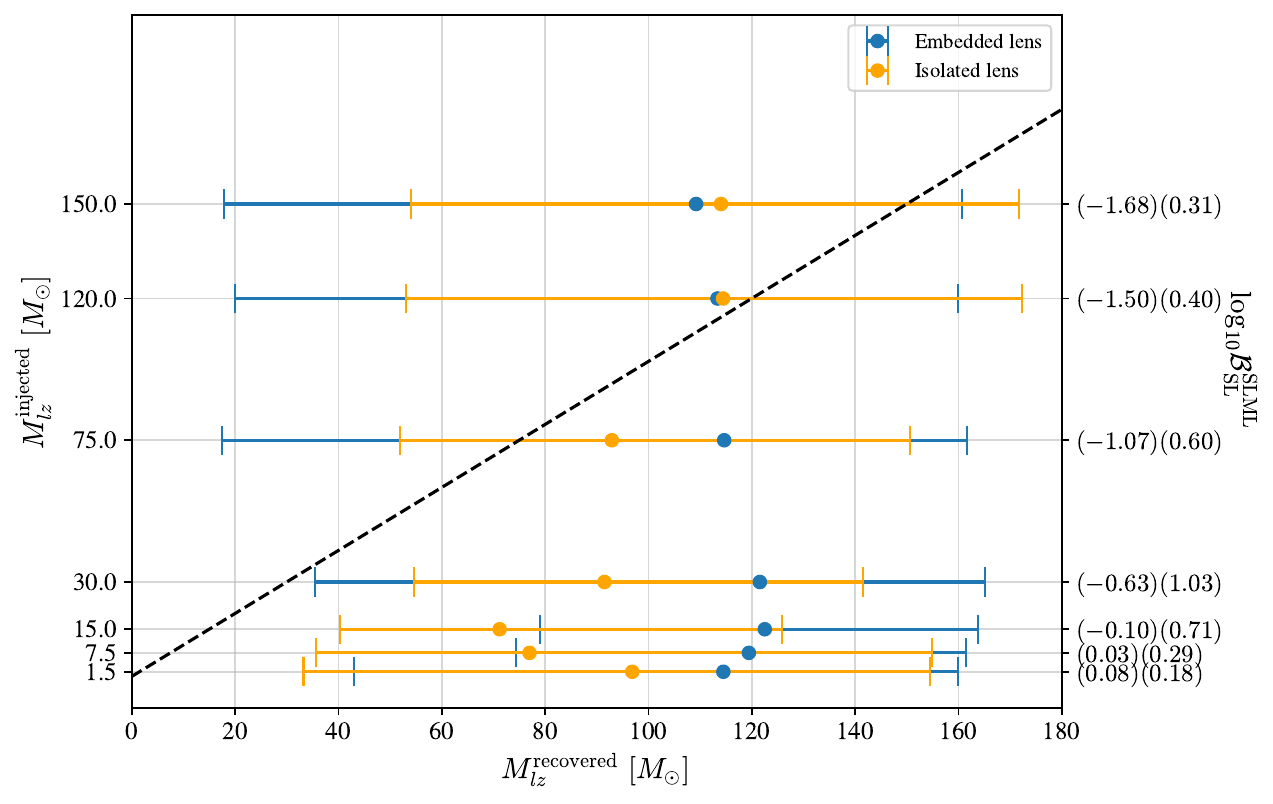}
        \caption{
        The recovered redshifted lens mass and the $\log_{10} \mathcal{B}_{\rm SL}^{\rm SLML}$ of injected signals with different injected microlens mass within the type-\Rom{1} embedded lens model and isolated lens model. 
        The black dashed line indicates that a recovered redshifted lens mass has the same value as injected one.
        The error bars indicate the 90 per cent confidence level.
        In the ineffective parameter estimation region of low mass injections, the $M_{lz}^{\rm rec}$ shows a flat trend as the uniform prior is recovered in such case.
        For the type-\Rom{1} embedded lens system, the $M_{lz}^{\rm rec}$ do not align the true value, indicating an ineffective parameter estimation.
        The $\log_{10} \mathcal{B}_{\rm SL}^{\rm SLML}$, the detectability of microlensing, is shown on the right axis, with that of type-\Rom{1} in the first parenthesis and that of type-\Rom{2} in the second.
        In general, the $\log_{10} \mathcal{B}_{\rm SL}^{\rm SLML}$ for type-\Rom{1} embedded lens are negative, showing that the method is incapable of detecting microlensing of the system.
        }
        \label{fig:combined_t1}
        
        \centering
        \includegraphics[height=6cm, width=9cm]{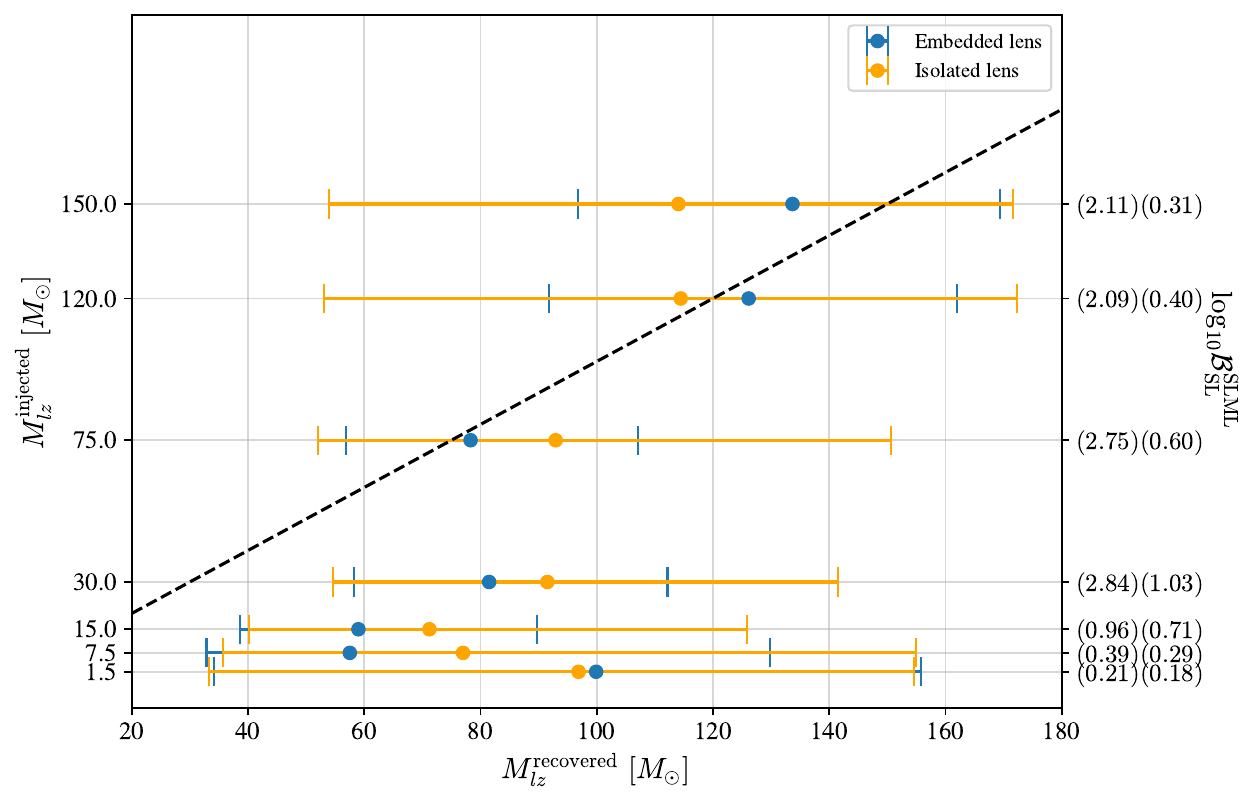}
        \caption{
        Same as \ref{fig:combined_t1}, but within the type-\Rom{2} embedded model.
        The isolated lens signal is independent of the image type, therefore showing the same trend as Fig. \ref{fig:combined_t1}. 
        The $M_{lz}^{\rm rec}$ for type-\Rom{2} embedded lens injections align the true value for higher $M_{lz}^{\rm inj}$. 
        The method could recover the microlensing details of a type-\Rom{2} microlensing scenario when the microlens has a higher mass.
        The $\log_{10} \mathcal{B}_{\rm SL}^{\rm SLML}$ is shown on the right axis.
        The $\log_{10} \mathcal{B}_{\rm SL}^{\rm SLML}$ are low for low mass injections. 
        For the type-\Rom{2} embedded lens signal, the $\log_{10} \mathcal{B}_{\rm SL}^{\rm SLML}$ are higher than 2 for a higher $M_{lz}^{\rm inj}$.
        It suggests that microlensing near type-\Rom{2} macroimage could be detected with high enough SNR.
        }
        \label{fig:combined_t2}
    \end{multicols}
\end{figure*}

Recently, \citet{seo2022improving} showed that microlensing of a type-\Rom{1} macroimage could be easier detected with the help of a secondary image, enabling a methodology to inspect tiny beating patterns otherwise hidden in noise.

The method assumed that the macromodel strongly lenses the source into two images, one of these images further microlensed by a star, while the other remains unaffected by any substructure within the macrolens.
In that case, we could in principle extract the source parameters from the unaffected image, so we can fix the source parameters including luminosity distance when we are conducting microlensing analysis on the (other) microlensed signal.
Then, the microlens's parameters such as its redshifted lens mass and impact parameter could be recovered more accurately, so as a strengthened confidence of microlensing detection.

As mentioned previously, when a microlens is embedded within a galaxy, it will lens the waveform in a way that differs non-trivially from that when the microlens is isolated.
Following \citet{seo2022improving}, we investigate the detectability of microlensing by these embedded lenses if we, as an approximation, use an isolated point mass lensing waveform as a template in Bayesian parameter estimation.

For the injected waveform, we use both type-\Rom{1} and type-\Rom{2} embedded lens systems with microlens masses $M_{lz}^{\rm inj} \in \{1.5, 7.5, 15, 30, 75, 120, 150\}$ to the right of the macroimage, as defined in Fig. \ref{fig:contours} (to the direction away from the source).
Note that the maximum mass is consistent with the analysis in previous sections($100 M_{\odot}$): since the lens is located at $z_{\rm l} = 0.5$, the redshifted lens mass would be $150 M_{\odot}$.
To compare the detectability using the isolated point mass template by \citet{seo2022improving}, we also inject waveforms lensed by the isolated point mass model as a fair test.
There is, therefore, a total of 21 injections with SNRs are fixed at 20.
For the macromodel, we use an SIS lens profile of $10^{10} {\rm M}_{\odot}$ with impact parameter $\eta_{\rm SIS} = 0.1 \theta_{\rm SIS}$ that produces macromagnifications of 11 and 9 for type-\Rom{1} and \Rom{2} images. 
The component masses of the BBH system are (30, 30) ${\rm M}_{\odot}$.
In the later part of the results, we also vary the macro-impact parameter to vary the macromagnifications of the images.

We use an isolated point mass lens waveform as a template. 
We follow \citet{seo2022improving} in which there is one strongly lensed signal being further microlensed, given two strongly lensed signals have been detected.
We quantify the detectability by $\log_{10} \mathcal{B}_{\rm SL}^{\rm SLML}$.
That is, we contrast two hypotheses: (i) that the signal is both strongly lensed and additionally microlensed ($\rm SLML$) versus (ii) that the waveform is only strongly lensed ($\rm SL$) and not microlensed.  

For comparison, we also perform an injection study where the injected waveform and the search template are both from the isolated lens model.
Notice that for the isolated lens model, because a microlens is added without the presence of the macromodel, the isolated lens model is independent of the image type.
Thus, the isolated lens waveforms used in detecting microlensing and for injections are the same set for both type-\Rom{1} and type-\Rom{2} microlensing systems.

\begin{figure*}
    \begin{multicols}{2}
        \centering
        \includegraphics[height=6cm, width=9cm]{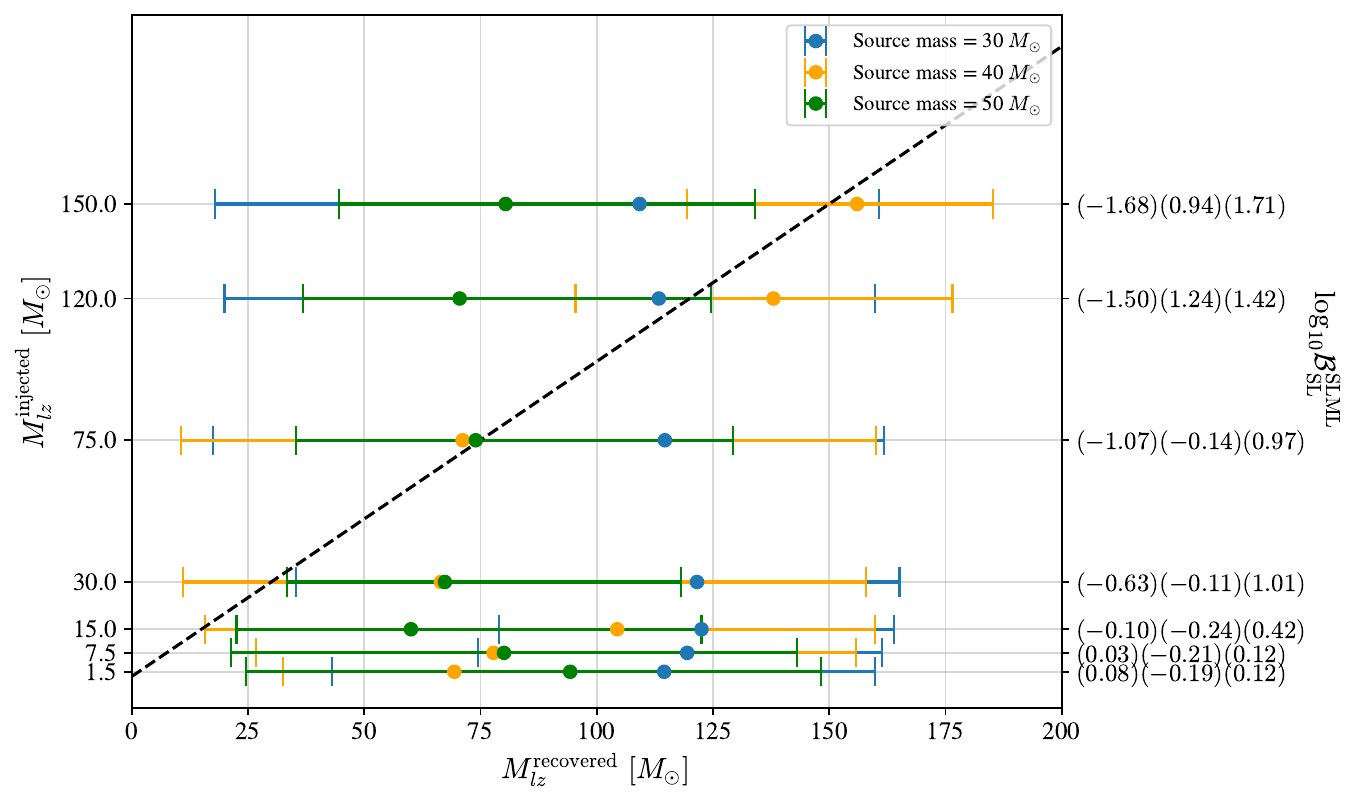}
        \caption{
        We vary the source component masses for type-\Rom{1} embedded lens.
        The chosen binary masses are $\{(30, 30), (40, 40), (50, 50)\} {\rm M}_{\odot}$.
        Same as \ref{fig:combined_t1}, the inferences of $M_{lz}^{\rm rec}$ are not correct.
        The $\log_{10} \mathcal{B}_{\rm SL}^{\rm SLML}$ is shown on the right axis, where the numbers in the parenthesis correspond to source masses $= 30, 40, 50 M_{\odot}$ respectively.
        In general, the $\log_{10} \mathcal{B}_{\rm SL}^{\rm SLML}$ is not showing any significant evidences of microlensing detection.
        }
        \label{fig:combined_dmt1}
        
        \centering
        \includegraphics[height=6cm, width=9cm]{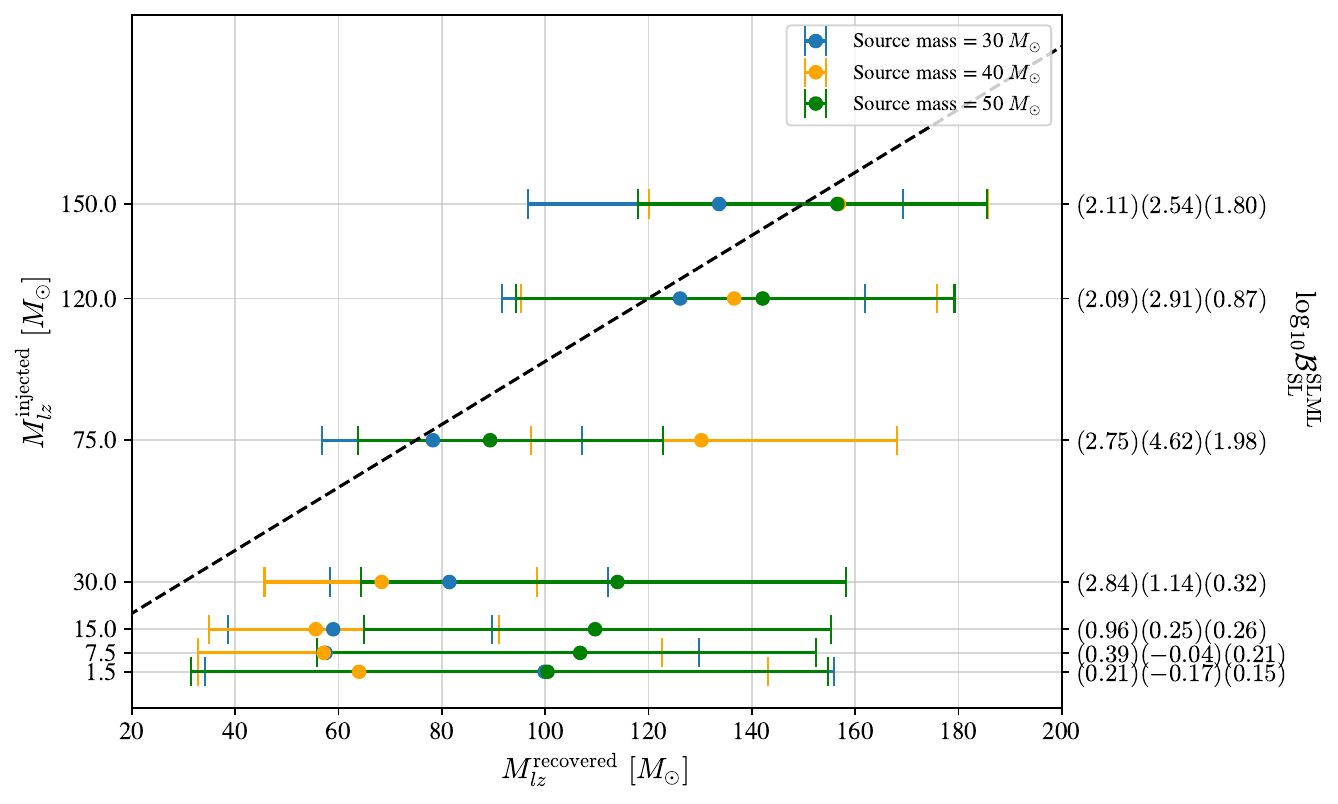}
        \caption{
        The type-\Rom{2} embedded lens model under variation of source component masses.
        The $M_{lz}^{\rm rec}$ align to the true value line for the higher injected microlens' masses.
        The $\log_{10} \mathcal{B}_{\rm SL}^{\rm SLML}$ is shown on the right axis, where the numbers in the parenthesis correspond to $\eta = 0.1, 0.2, 0.3 \theta_{\rm SIS}$ respectively.
        type-\Rom{2} microlensing shows positive Bayes factor due to waveform similarity with isolated point mass's template.
        }
        \label{fig:combined_dmt2}

        \centering
        \includegraphics[height=6cm, width=9cm]{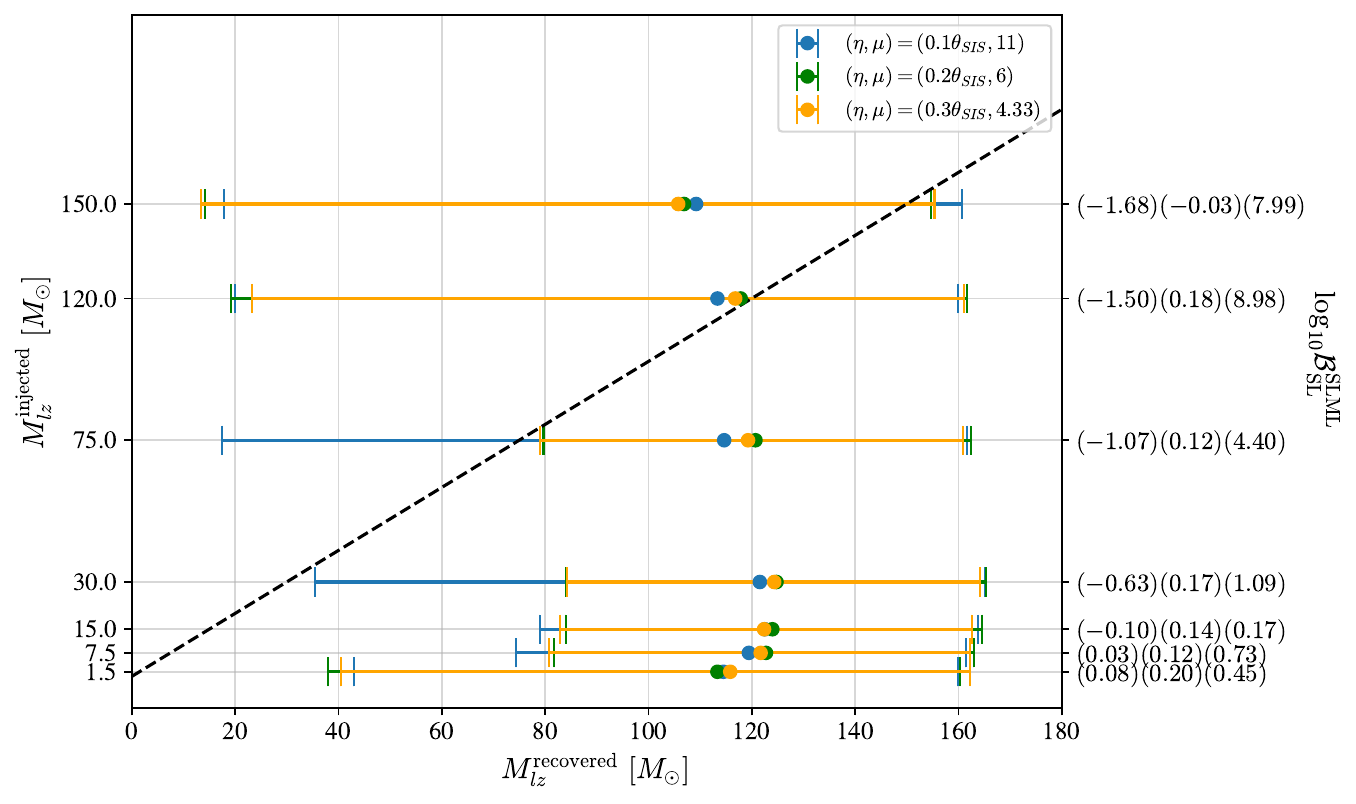}
        \caption{
        We change the macromagnification of the macroimage from type-\Rom{1} embedded lens model.
        The $M_{lz}^{\rm rec}$ are flat and do not show any trend. The method is unable to recover the microlens mass for type-\Rom{1} embedded lens system.  
        As the macromagnification being smaller, the influence of macrolens is reduced, the $\log_{10} \mathcal{B}_{\rm SL}^{\rm SLML}$ are higher as well, ignoring the ineffective result of low mass injection.
        }
        \label{fig:combined_dyt1}
        
        \centering
        \includegraphics[height=6cm, width=9cm]{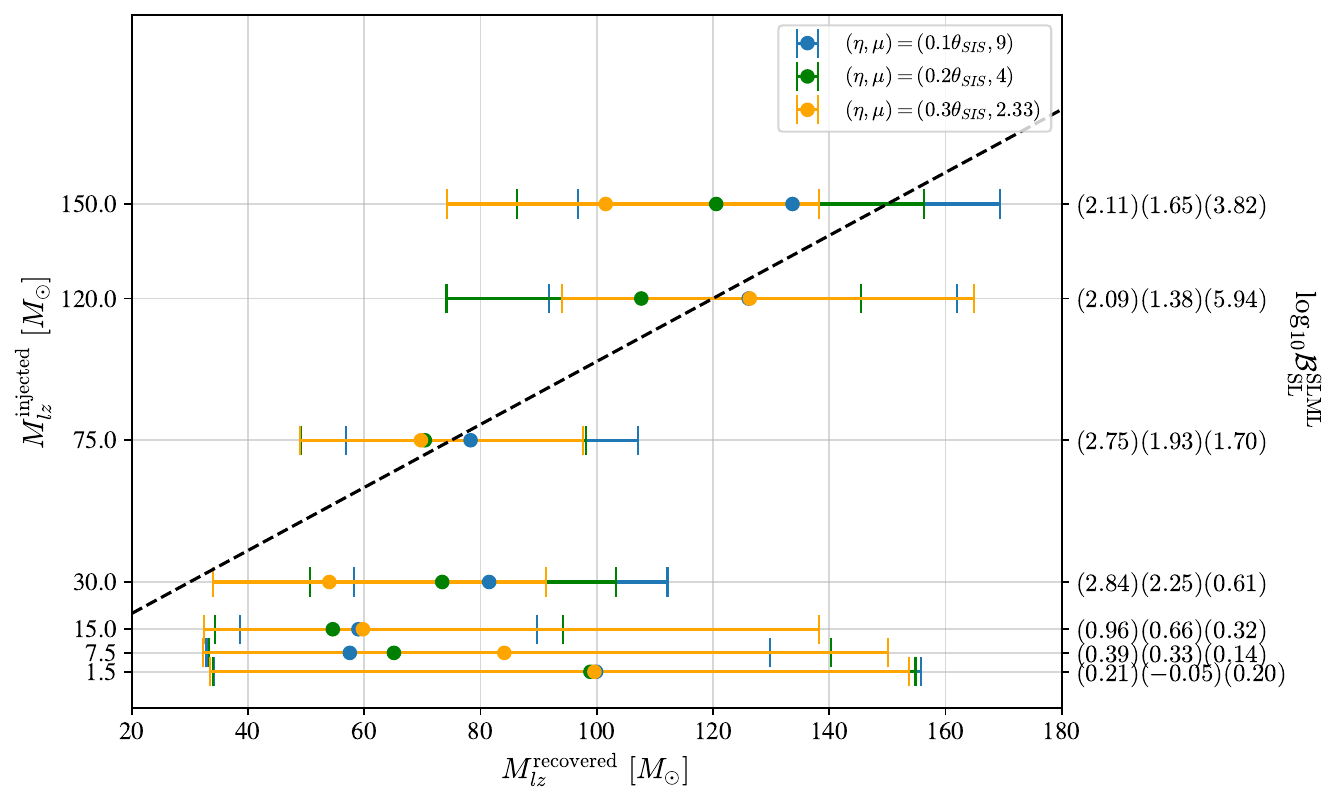}
        \caption{
        Same as \ref{fig:combined_dyt1}, but for type-\Rom{2} embedded lens.
        In higher mass region, the recovered redshifted massses of different models are generally closer to the injected value, compared to type-\Rom{1}'s case.
        Same as Fig. \ref{fig:combined_dyt1}, as the influence of the microlens reduces, the $\log_{10} \mathcal{B}_{\rm SL}^{\rm SLML}$ are higher.
        }
        \label{fig:combined_dyt2}
    \end{multicols}     
\end{figure*}

Figs \ref{fig:combined_t1} and Figure \ref{fig:combined_t2} show the recovered redshifted mass $M_{lz}^{\rm rec}$ from parameter estimation for type-\Rom{1} and type-\Rom{2} embedded lens systems respectively, along with $M_{lz}^{\rm rec}$ for an injected isolated lens system for comparison.
For all injected lens systems, the $M_{lz}^{\rm rec}$ are not recovered well for low mass injections.
As the wave optics would be more prominent for low mass microlens, the amplification factor is suppressed and less of the beating pattern is seen, thus the recovered mass is imprecise even when the template and the waveform are from the same isolated lens model.
In region that the $M_{lz}^{\rm rec}$ does not align to the true value, the parameter inference is recovering the uniform prior, indicating ineffective parameter estimations.
The recovered posterior of $M_{lz}^{\rm rec}$ for the type-\Rom{2} microlensing system with higher $M_{lz}^{\rm inj}$ is more informative, centring around the injected value, instead of returning the flat prior.
Thus, the isolated lens template is able to reconstruct approximately the microlensed waveform near a type-\Rom{2} macroimage.
On the other hand, the $M_{lz}^{\rm rec}$ for the type-\Rom{1} embedded lens systems do not align the true value, but rather follow the priors.
This means that the isolated lens template could not reconstruct the microlensing details of the more complicated type-\Rom{1} embedded lens system.
It suggests the need of a more advanced waveform as template to recover the true microlens's mass in a realistic scenario.

\begin{figure}
    \centering
    \includegraphics[height=6cm, width=8cm]{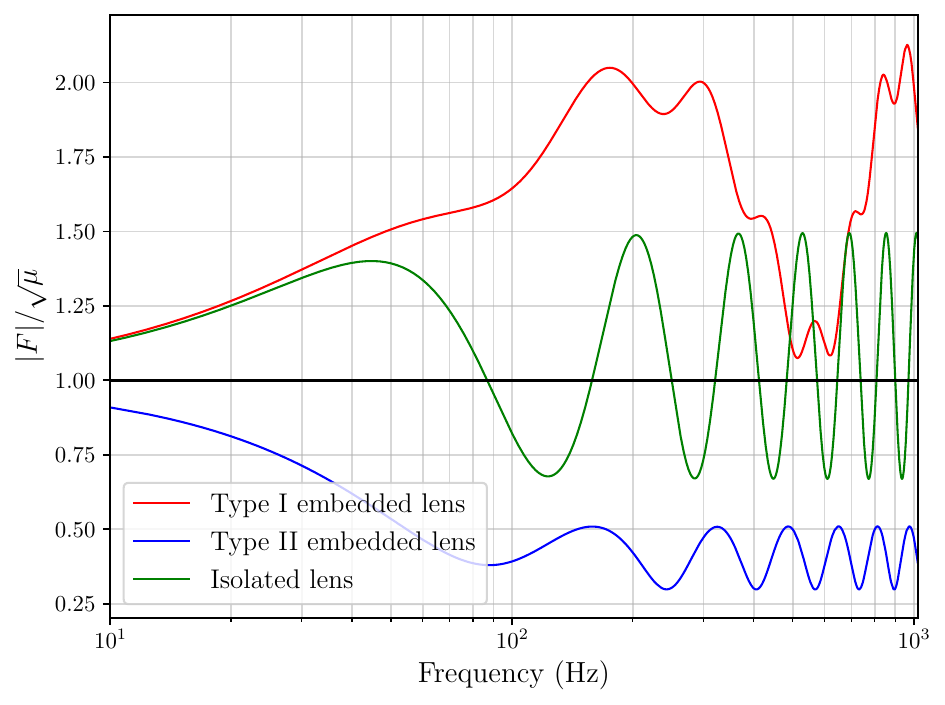}
    \caption{
    The amplification factors of microlensing near type-\Rom{1}, type-\Rom{2} macroimages as embedded lens systems and isolated lens model. The microlens with mass $m = 100 {\rm M}_{\odot}$ is placed $1 \theta_m$ away from the right-hand side of the macroimage. The waveform morphology of isolated lens system and the type-\Rom{2} embedded lens system are similar due to the same number of images.}
    \label{fig:different_models}
\end{figure}

The right axis of Figs \ref{fig:combined_t1} and \ref{fig:combined_t2} show the $\log_{10} \mathcal{B}_{\rm SL}^{\rm SLML}$ for the type-\Rom{1} and type-\Rom{2} embedded lens and isolated lens systems.
Generally, the Bayes factor for the injected type-\Rom{1} images are negative and hence there is no evidence of detecting the type-\Rom{1} microlensing.
On the other hand, the results shown in Fig. \ref{fig:combined_t2} shows the microlensing signature of type-\Rom{2} images could be detected for more massive microlenses.
For the cases of $M_{lz}^{\rm rec} > 30 {\rm M}_{\odot}$, the $\log_{10} \mathcal{B}_{\rm SL}^{\rm SLML}$ goes beyond 2, showing positive evidence of detecting microlensing.
As shown in Fig.~\ref{fig:variation_mass} the mass of the microlens gets larger, the beating pattern would also be more significant and the microlensing would be easier to search with.
The differences in detecting microlensing of type-\Rom{1} and type-\Rom{2} systems with isolated lens template could be explained by the number of images produced.
In typical type-\Rom{1} microlensing scenarios \citep{Cheung_2021}, there would be four microimages produced.
While for the type-\Rom{2} case, there are typically two microimages instead, which coincides with the number of images in the isolated lens model.
Therefore, type-\Rom{2} microlensing amplifcation factor would share a similar waveform morphology as isolated lens's (Fig. \ref{fig:different_models}), resulting in a higher Bayes factor for the type-\Rom{2} microlensing case.

To test whether the above results hold in general, we repeat the same analysis, but with different source parameters for the BBH merger and different macromodel parameters (i.e. $\eta_{\rm SIS}$, hence changing the macromagnification of the macroimage). 
We vary the component masses of the source BBH system, and thus the frequency evolution of the waveforms are different from the previous case.
The varied component masses are $\{(30, 30), (40, 40), (50, 50)\} {\rm M}_{\odot}$.
The macromagnification are varied by changing the macro-impact parameters $\eta_{SIS}$ of the GW.   
\begin{equation}
\centering
    \mu_{\pm} = \pm 1 + 1/\eta_{SIS}\,
\end{equation}
where $\mu_{\pm}$ is the macromagnification of type-\Rom{1} or type-\Rom{2} macroimage.
As the macro-impact parameter is smaller, the GW is under a smaller influence of the macrolens, leading to a smaller macromagnification.
The varied macromagnifications are $\{(11, 9), (6, 4), (4.33, 2.33)\}$ for type-\Rom{1} and type-\Rom{2} macroimages respectively.
For each set of the component mass and macromagnification, we inject $\{1.5, 7.5, 15, 30, 75, 120, 150\} {\rm M}_{\odot}$.

Figs \ref{fig:combined_dmt1}, \ref{fig:combined_dmt2}, \ref{fig:combined_dyt1} and \ref{fig:combined_dyt2} show the recovered redshifted lens mass $M_{lz}^{\rm rec}$ versus the source parameters and macromagnifications for type-\Rom{1} and type-\Rom{2} embedded lens systems.
The type-\Rom{1} and type-\Rom{2} cases show similar trends as Figs \ref{fig:combined_t1} and \ref{fig:combined_t2} respectively.
For type-\Rom{1} embedded lens system, $M_{lz}^{\rm rec}$ can be better recovered when the source component masses is $40 {\rm M}_{\odot}$. 
This is because the luminosity distance of the signal is better recovered from the parameter estimation of the strongly lensed signal:
the recovered redshifted lens mass and impact parameter will be biased to compensate for the change in the signal amplitude if the luminosity distance is not well-recovered.
The injected lens masses in higher mass region could be recovered within the 90 $\%$ confidence region for type-\Rom{2} embedded lens system.

Fig. \ref{fig:combined_dmt1} shows $\log_{10} \mathcal{B}_{\rm SL}^{\rm SLML}$ for the type-\Rom{1} embedded lens system under variation of source parameter. 
When the component masses of the binary system increase, the chirp frequency of the GW signal becomes lower.
The effect of amplification factor at low frequency region would thus be more prominent. 
As shown in Fig. \ref{fig:different_models}, the amplification factors of embedded lens model and isolated point mass model both converge to unity at low frequencies. 
The difference between the two decreases as the the frequency goes down, thus the isolated point mass model template is being more capable to capture the embedded lens model signal.
Moreover, for source of higher binary masses, the signal is louder and that in turn increases the Bayes factor of microlensing detection.
Therefore, $\log_{10} \mathcal{B}_{\rm SL}^{\rm SLML}$ is expected to be higher as the source's component masses increase.
Fig. \ref{fig:combined_dyt1} shows $\log_{10} \mathcal{B}_{\rm SL}^{\rm SLML}$ for the same system when the macromagnification is varied.
We see that $\log_{10} \mathcal{B}_{\rm SL}^{\rm SLML}$ is significantly higher when the macromagnification is low.
As the impact of the macrolens becomes smaller, the amplification factor is closer to that of the isolated model.
Therefore, the method could detect microlensing in such case with a greater confidence.

For type-\Rom{2} embedded lens system under variation of the source BBH parameter, as shown by Fig. \ref{fig:combined_dmt2}, $\log_{10} \mathcal{B}_{\rm SL}^{\rm SLML}$ is generally positive due to the waveform similarity between embedded lens model and point mass model, as discussed above. 
Fig. \ref{fig:combined_dyt2} shows the $\log_{10} \mathcal{B}_{\rm SL}^{\rm SLML}$ for the type-\Rom{2} embedded lens system with changing macromagnification, showing similar trends as Fig. \ref{fig:combined_t2}.
Not only is the amplification factor of type-\Rom{2} embedded lens system closer to that of isolated model, but the reduced influence of the macromodel also increases the detectability of the signal.
Among all the cases, the type-\Rom{2} embedded lens system gives the highest Bayes factor when an isolated point mass lens template is used.

\section{Discussion}
\label{sec:implication}

Similarly to \citet{Diego:2019lcd} and \cite{mishra}, we generally find that there would be a demagnification in amplitude for a microlensed type-\Rom{2} macroimage.  
This demagnification reduces our ability to detect microlensed events near type-\Rom{2} macroimages. 
On the other hand, the effects of microlensing on the waveform tends to be \emph{greater} for type-\Rom{2} images, as can be seen, for example, when comparing Fig. \ref{fig:mismatch} with fig. 9 of \cite{Cheung_2021}. 
Therefore, the detectability will likely be a tug-of-war between the decrease in the SNR due to de-magnification and the increase in the waveform mismatch. 
At this moment, it is unclear which effect wins out. 

As mentioned in the previous section, many strong lensing science cases, such as the localization of the host galaxies of merging black holes as well as lensing cosmography~\citep{hannuksela2020localizing} rely on obtaining accurate magnification measurements from the macroimages, unhindered by microlensing (see also~\citep{Oguri:2010a,Cheung_2021}, for discussion).
Fig.~\ref{fig:mismatch} shows that for microlenses of mass $m \lesssim 10 M_\odot$, the magnification of the image would not significantly affect these studies if the targeted accuracy is $\sim$ 10 per cent. 
The results seem independent of the impact parameter of the microlens. 
Similarly to the results in \cite{Cheung_2021,mishra}, we can be confident that the overall magnifications between images of both type-\Rom{1} and \Rom{2} are not significantly altered by stellar-mass microlensing by low-mass stars, except in rare cases where more massive microlenses are present.

As mentioned in \ref{sec:methods}, the window size for type-\Rom{2} macroimage is significantly larger, giving rise to a high computational cost.
Recently, \citet{Shan-LensingBoundaries} has proposed a method to reduce the size of the integration boundary, which could improve the computational efficiency of the amplification factor significantly.

In summary, we have analysed the amplification factor for a microlensed type-\Rom{2} macroimage. 
Our previous research studied the effects of microlenses on strongly-lensed type-\Rom{1} images. 
Here, we have completed the picture by extending our previous research to type-\Rom{2} strongly-lensed images. 
In line with \citet{Diego:2019lcd,mishra}, we find the wave optics suppresses the amplification factor in low frequency and demagnification is expected for the microlensed type-\Rom{2} images.
Furthermore, we find that the mismatch for type-\Rom{2} images is greater than it is for type-\Rom{1} images.
The wave optics analysis allows an accurate treatment of the amplification of the GW lensed images. 
We expect future work to further study fields of microlenses.

On the detection side, the analysis techniques combining strong lensing with microlensing by \citet{seo2022improving} could search the microlensing signature from a microlensed type-\Rom{2} macroimage only with sufficient SNR and high microlens's mass.
Meanwhile, as the component masses of the source becomes higher and the impact of the macrolens reduces, the confidence of microlensing detection increases even for a microlensed type-\Rom{1} macroimage.
Microlensing is not distinguishable for sources with low component masses and systems heavily influenced by the macromodel (i.e. low macro impact parameter) with Bayesian inference using isolated point mass model.
On the recovery of microlens's parameters, the redshifted lens mass of the microlens in such cases could be recovered for a microlensed type-\Rom{2} macroimages with high microlens's mass.
The method is unable to recover the microlens's parameter when the microlens's mass is low and for type-\Rom{1} embedded lens systems generally.
While a point mass lens model could be used as an effective template for detection type-\Rom{2} microlensed images, in the sense that it returns a positive Bayes factor and an informative posterior consistent with the injection, a more accurate way to model microlensing affects will be crucial for confident detection and parameter estimation on generic microlensed GWs.
Our results are valid for low macromagnification regime ($\mu \lesssim 10$), future work of parameter estimation could be conducted beyond the limit of low magnification.

\section*{Acknowledgements}
We thank Anuj Mishra, Anupreeta More, Alvin Ka-Yue Li, Rico Ka-Lok Lo,  Soumen Roy, and Miguel Zumalac\'{a}rregui
for useful comments and discussion. 
M.H.Y.C. is supported by NSF Grants No. AST-2006538, No. PHY-2207502, No. PHY-090003, and No. PHY-20043, and NASA Grants No. 19-ATP19-0051, No. 20-LPS20-0011 and No. 21-ATP21-0010, the Croucher Foundation of Hong Kong.
O.A.H. and T.G.F.L are partially supported by grants from the Research Grants Council of the Hong Kong (Project No. 24304317 and 14306419), The Croucher Foundation of Hong Kong and Research Committee of the Chinese University of Hong Kong.
The authors acknowledge the Texas Advanced Computing Center (TACC) at The University of Texas at Austin for providing HPC resources that have contributed to the research results reported within this paper. URL: \url{http://www.tacc.utexas.edu}.

\section*{Data Availability}
The data underlying this article will be shared on reasonable request
to the corresponding authors.

\bibliographystyle{mnras}
\bibliography{microlensing} %

\section*{Appendix: Microlenses placed at different angles from the macroimage}

\begin{figure}
    \centering
    \includegraphics[height=6cm, width=8cm]{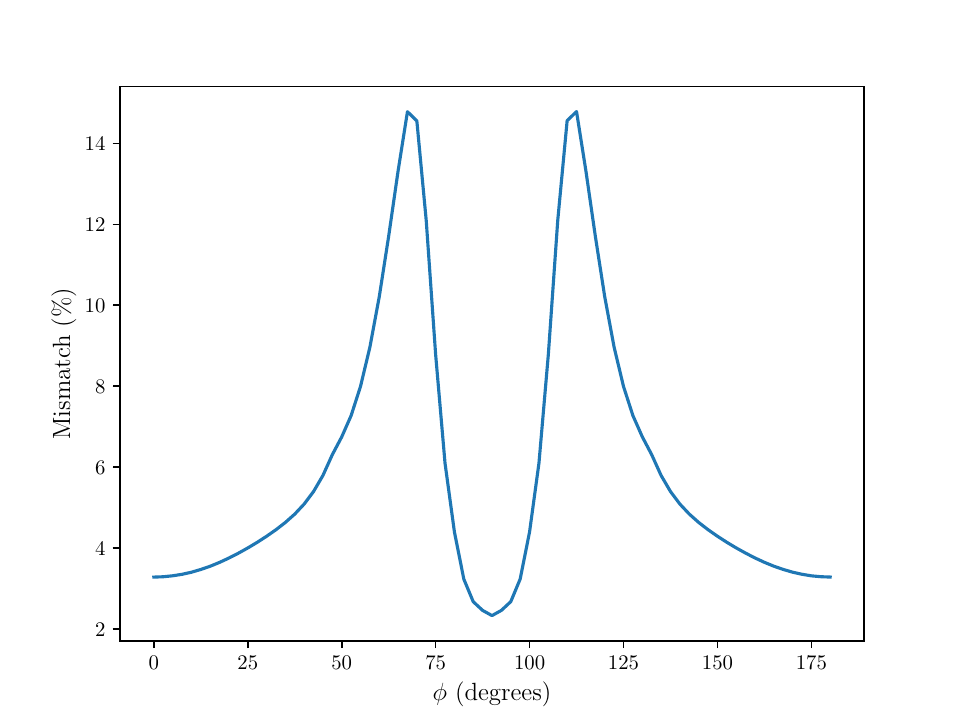}
    \caption{The mismatch of type-\Rom{2} $h_{\rm L, full}$ and $h_{\rm L, macro}$ in the parameter space of the relative angle $\phi$. 
    By the topology of the saddle point contours shown in Fig. \ref{fig:contours}, the results is symmetric about $\phi = 180 ^{\circ}$.
    The mismatch is greater around the ridges of the contours.}
    \label{fig:mismatch_angle}
\end{figure}

We have shown that the mismatch between the microlensed and un-microlensed waveforms is independent of the micro impact parameter when the microlens mass is small.
In this appendix, we will show that the mismatch varies rather significantly if we place the microlens to a different angular direction from the macroimage for type-\Rom{2} images.
Fig. \ref{fig:mismatch_angle} shows the mismatch as a function of $\phi$ between 0 $^{\circ}$ and 180 $^{\circ}$, where $\phi$ is the relative angle from the SIS center to the microlens measured at the macroimage (the results for $\phi \in (180^{\circ}, 360^{\circ})$ can be obtained trivially by symmetry). 
The macromodel is the same as the Section \ref{sec:results}.
We place the microlens of $100 {\rm M}_{\odot}$ at $\eta_{m} = 1 \theta_{m}$.
The binary masses are (30, 30) ${\rm M}_{\odot}$.  
Contrary to \citep{Cheung_2021}, the relative angle plays a more important role in type-\Rom{2} microlensing due to the topology of the time delay contours around a type-\Rom{2} image shown in \ref{fig:contours}:
The mismatch is higher when the lens is placed around the ridges of the contour. 
Even when the lens is placed at $0^{\circ}$ which shows the lowest mismatch, the microlensed waveform still significantly deviates from the unlensed waveform for mismatch >3 per cent.
Therefore, regardless of the relative angle, microlensing of type-\Rom{2} macroimages would still be detectable given sufficient SNR.

\bsp	%
\label{lastpage}
\end{document}